\documentclass{article}

\usepackage{PRIMEarxiv}
\usepackage[utf8]{inputenc} % allow utf-8 input
\usepackage[T1]{fontenc}    % use 8-bit T1 fonts
\usepackage{hyperref}       % hyperlinks
\usepackage{url}            % simple URL typesetting
\usepackage{booktabs}       % professional-quality tables
\usepackage{amsfonts, amsmath}       % blackboard math symbols
\usepackage{nicefrac}       % compact symbols for 1/2, etc.
\usepackage{microtype}      % microtypography
\usepackage{lipsum}
\usepackage{fancyhdr}       % header
\usepackage{graphicx}       % graphics
\graphicspath{{media/}}     % organize your images and other figures under media/ folder
\usepackage{epstopdf}% To incorporate .eps illustrations using PDFLaTeX, etc.

\title{Generalized statistics: applications to data inverse problems with outlier-resistance
}

\author{
  João V. T. de Lima  \\
  Department of Theoretical and Experimental Physics \\
  Federal University of Rio Grande do Norte \\
  Natal, RN, 59078-970, Brazil.\\
  \texttt{tomazvictor13@gmail.com} \\
   \And
  Sérgio Luiz E. F. da Silva  \\
  Seismic Inversion and Imaging Group \\
  Fluminense Federal University \\
  Niterói, RJ, 24210-346, Brazil\\
  \texttt{sergioluizsilva@id.uff.br} \\
   \And
  João M. de Araújo  \\
  Department of Theoretical and Experimental Physics \\
  Federal University of Rio Grande do Norte \\
  Natal, RN, 59078-970, Brazil.\\
  \texttt{joaomedeiros@fisica.ufrn.br} \\
   \And
  Gilberto Corso \\
  Department of Biophysics and Pharmacology \\
  Federal University of Rio Grande do Norte \\
  Natal, RN, 59078-970, Brazil.\\
  \texttt{gfcorso@gmail.com} \\
   \And
  Gustavo Z. dos Santos Lima \\
  School of Science and Technology \\
  Federal University of Rio Grande do Norte \\
  Natal, RN, 59078-970, Brazil.\\
  \texttt{guzampier76@gmail.com} \\
}

\begin{document}
\maketitle

\begin{abstract}
The conventional approach to data-driven inversion framework is based on Gaussian statistics that presents serious difficulties, especially in the presence of outliers in the measurements. In this work, we present maximum likelihood estimators associated with generalized Gaussian distributions in the context of Rényi, Tsallis and Kaniadakis statistics. In this regard, we analytically analyse the outlier-resistance of each proposal through the so-called influence function. In this way, we formulate inverse problems by constructing objective functions linked to the maximum likelihood estimators. To demonstrate the robustness of the generalized methodologies, we consider an important geophysical inverse problem with high noisy data with spikes. The results reveal that the best data inversion performance occurs when the entropic index from each generalized statistic is associated with objective functions proportional to the inverse of the error amplitude. We argue that in such a limit the three approaches are resistant to outliers and are also equivalent, which suggests a lower computational cost for the inversion process due to the reduction of numerical simulations to be performed and the fast convergence of the optimization process.
\end{abstract}

% keywords can be removed

\section{Introduction}\label{sec1}
\label{sec:introduction}
The estimation of physical model parameters  from observed data is a frequent problem in many areas, such as in machine learning \cite{RAISSI2019686,HARLIM2021109922_2021}, geophysics \cite{artigomestrado2017,FWI_EAGEITALY_PLOSONE2020}, biology \cite{InverseProblemsBiologia1,InverseProblemsBiologia2}, physics \cite{BA2018300,Razavy}, among others \cite{inverseproblems_quimica,plosonewilton_2019,HEIDEL2021109865}. Such a task is solved through the so-called inverse problem, which consists of identifying physical parameters that can not be directly measured from the observations \cite{tarantola2005book}. From a practical viewpoint, in the inverse problem, physical model parameters are estimated by matching the calculated data to the observed data by optimizing an objective function \cite{MENKE20121}. The objective function in the least-squares sense is widely used, which is based on the assumption that errors are independent and identically distributed (\textit{iid}) by a standard Gaussian probability distribution \cite{tarantola2005book}. Although this approach is quite popular, the least-squares estimation is biased if the errors are non-Gaussian, violating the Gauss-Markov theorem \cite{gauss_markovtheorem_KENDALL,THOMSON2014219}. Indeed, just a few outliers are enough for the least-squares criterion to be inappropriate \cite{robustMethods1988}.

In this way, a lot of non-Gaussian criteria have been proposed to mitigate the inverse problem sensitivity to aberrant measurements (outliers). The most common criterion to deal with non-Gaussian errors is based on the $L_1$-norm of the difference between the calculated and the observed data, in which the errors are assumed to be \textit{iid} according to a double exponential distribution (Laplace distribution) \cite{tarantolanormal1_1987}. Although this approach is known for being outlier-insensitive, this criterion is singular in cases where the error is null (or close to zero). Thus, it is necessary to assume that the absolute error is greater than zero according to the machine precision used, which generates an indeterminacy problem from a computational point of view. To avoid the singularity of this approach, hybrid criteria which combine the least-squares distance with the least-absolute-values criterion ($L_1$-norm) have been proposed and successfully applied for parameter robust estimation \cite{huberOriginal1973,HuberFWI_Guitton,BubeLangan_1997_hybridObjectiveFunction}. However, hybrid approaches require the determination of a threshold parameter, which demands boring trial-and-error investigations, increasing the computational cost \cite{residualnormbrossier}.

Indeed, objective functions based on heavy-tailed probability distributions, such as the Cauchy-Lorentz distribution \cite{CauchyRegression_2017} and the Student's \textit{t}-distribution \cite{Ubaidillah_2017_studentT}, have demonstrated robust properties for unbiased data inversion. However, both approaches assume a fixed probability distribution of errors, not adapting to the particularities of the model or data at hand. In this sense, objective functions based on generalized distributions are interesting because they might be adapted to the specificities of the erratic data by selecting an adequate free-parameter. In fact, several generalized approaches have been proposed to deal with erratic data \cite{GeneralizedGaussian_SEG2015,Sacchi_SEG_generalized_2020,PhysRevE.104.024107,PSI_Jackson_2021,FWI_qLaplace_2021}. Thus, generalized distributions based on the Rényi, Tsallis and Kaniadakis statistics have generated objective functions robust to erratic noise \cite{sergio2021extensive}. 

\begin{figure*}[!htb]
    %\resizebox{\textwidth}{!}{\includegraphics{fig_01}}
    \centering
    \includegraphics[width=14cm]{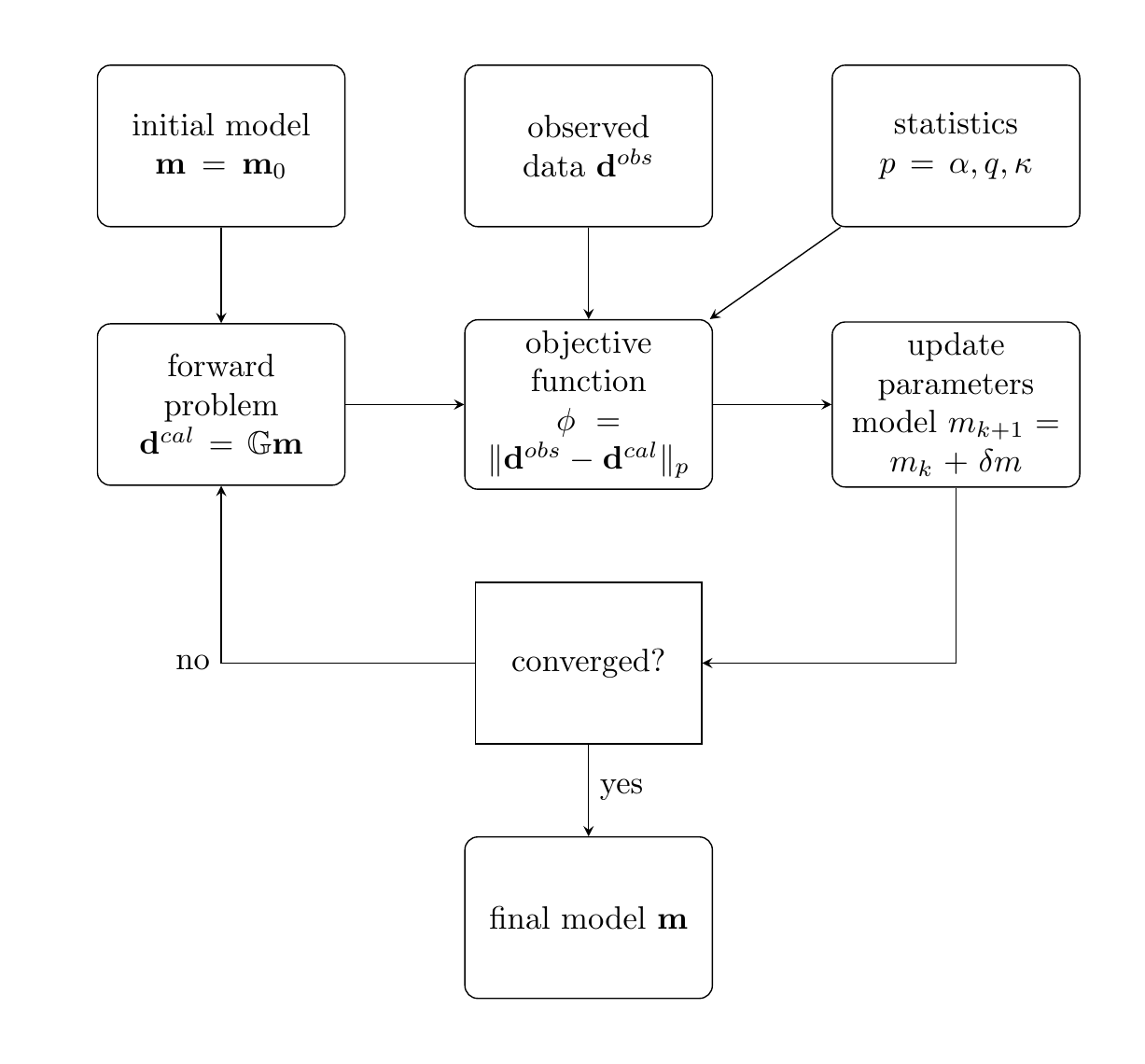}
    \caption{Workflow of the computational experiments.}
    \label{fig:diagrama}
\end{figure*}

In this work, we consider deformed Gaussian distributions associated to generalized statistical mechanics in the sense of Rényi ($\alpha$-statistics) \cite{Renyi1965informationTheory}, Tsallis ($q$-statistics) \cite{Tsallis1988} and Kaniadakis ($\kappa$-statistics) \cite{kaniadakis2001} to mitigate the undesirable effects of outliers on estimates of physical parameters. In particular, we place the objective functions based on $\alpha$-, $q$- and $\kappa$-generalizations in the broad context of the Gauss' law of error \cite{Tanaka2019,errorLawTsallisSuyari,WADA200689}, see Refs.~\cite{sergio2021extensive,daSilva2020robust,PhysRevE.101.053311}. The three deformed Gaussian distributions mentioned above have already demonstrated robust properties in many applications \cite{sergio2021extensive,daSilva2020robust,PhysRevE.101.053311,igoTsallisEntropy,FWIqGAuss_EAGE_2020,Lima2021,daSilva_2021_EPJPLUS_q_k_Laplace,daSilva_SEG_IMAGE_2021_qGaussian}. However, the entropic index associated with each of these approaches that make the inversion process more robust requires thorough investigation. In this regard,  we analyse and compare the generalized objective functions from a statistical and numerical point of view in order to obtain the optimum value of the entropic index. The workflow of the experiments employed in this work is summarized in Fig.~\ref{fig:diagrama} in which it is represented a flowchart of an inverse problem. We call attention that in our framework, generalized statistics define the norm employed in the inversion problem solution.

We have organized this article as follows. In Section \ref{sec:generalizedStatisticsSeismic} we present a brief review on the solution of inverse problems using the maximum likelihood method in the conventional framework, as well as in the framework of Rényi, Tsallis and Kaniadakis. Moreover, in Section \ref{sec:comparacao} we discuss the similarities among the generalized objective functions by considering a numerical test whose purpose is to estimate line model parameters; and finally, Section \ref{sec:numericalExperiment} is destined to apply the methodology presented in the article to address a classic geophysical problem that consists of estimating the acoustic impedance model using seismic data post-stack contaminated with spike noise.

\section{A brief review of generalized statistical in inverse theory: Maximum likelihood methods}\label{sec:generalizedStatisticsSeismic}

An inverse problem is formulated as an optimization task, in which an objective function describes how well the estimated model matches the measurements \cite{tarantola2005book}. In this regard, the model parameters $\mathbf{m}$ are obtained by solving the following linear system \cite{tarantola2005book, MENKE20121}:
\begin{equation}
    \mathbf{d}^{cal} = \mathbb{G} \mathbf{m} ,
    \label{eq:linear-model}
\end{equation}
where $\mathbf{d}^{cal}$ are the calculated data, in which $\mathbb{G}$ represents the forward operator. 
It is worth emphasizing that inverse problems are ill-posed \cite{hadamard}, which means that the solution of Eq.~\eqref{eq:linear-model} is not unique. This is due, in general, to the fact that the observed data are band-limited and noisy. In this way, it is necessary to find methodologies capable to solve Eq.~\eqref{eq:linear-model} for cases where the observed data, $\mathbf{d}^{obs}$, is corrupted by measurement errors, limitations in data acquisition, among other factors.

In the conventional approach, model parameters are estimated by maximizing the objective function is derived from the maximization of the Boltzmann-Gibbs-Shannon entropy (BGS):
\begin{equation}
    \mathcal{S}_{BGS}\big[ p \big] = -\sum_{i=1}^N p(x_i) \, \ln{\Big( p(x_i) \Big)} \, ,
    \label{eq:shannon-entropy}
\end{equation}
subject to the following constraints:
\begin{equation}
    \sum_{i=1}^N p(x_i) = 1 \quad \textrm{(normalization condition),}
    \label{eq:vinculos_norm}
\end{equation}

\begin{equation}
    \sum_{i=1}^N x_i^2 \, p(_ix)  = 1 \quad \textrm{(unity variance),}
    \label{eq:vinculos_var}
\end{equation}
where $p$ is a probability function and $\mathbf{x} = \mathbf{d}^{obs} - \mathbf{d}^{cal} = \{x_1, x_2, ..., x_N\}$ represents the difference between observed and calculated data. As well known in the literature, the probability distribution determined from the optimization of the BGS functional entropy subject to the constraints in Eqs.~\eqref{eq:vinculos_norm} and \eqref{eq:vinculos_var} corresponds to the standard Gaussian distribution (see, for instance, Section 2 of Ref.~\cite{Lima2021}):
\begin{equation}
    p(x) = \frac{1}{\sqrt{2\pi}}\exp{\left( -\frac{1}{2}x^2  \right)}.%\textcolor{blue}{.}
    \label{eq:gaussian-PDF}
\end{equation}

In other words, inverse problems, in the conventional framework, are solved from the premise that errors are independent and identically distributed (iid) according to a standard Gaussian likelihood function, which is formulated as the following optimization problem:
\begin{equation}
    \begin{split}
        \max_\mathbf{m} \mathcal{L}_G (\mathbf{m}) &= \prod _{i = 1}^{N} p\Big(x_i (\mathbf{m})\Big) \\
        &= \left( \frac{1}{\sqrt{2\pi}} \right)^N \exp{\left( -\frac{1}{2}\sum_{i=1}^N x_i^2(\mathbf{m}) \right)}.
    \end{split}
    \label{eq:gaussian-likelihood}
\end{equation}
A likelihood function is a useful tool to estimate physical parameters from the empirical data. In practice, inverse problems based on the Gauss' law of error are formulated in a least-squares sense. To see this, we notice that maximizing the likelihood function in \eqref{eq:gaussian-likelihood} is equivalent to minimizing the negative of the log-likelihood:
\begin{equation}
    \begin{split}
        \min_\mathbf{m} \phi (\mathbf{m}) &= \frac{1}{2}\sum_{i=1}^N x_i ^2 (\mathbf{m})\\
        &= \frac{1}{2} \sum_{i=1}^N \big( d^{obs}_i - {d}^{cal}_i (\mathbf{m}) \big) ^2.
    \end{split}
    \label{eq:classical-misfit}
\end{equation}
 
However, it is worth emphasizing that in several problems the errors are non-Gaussian and, therefore, in such contexts the conventional approach becomes inefficient, especially in the presence of aberrant measures (outliers) \cite{Claerbout1973}.

As is already known, we obtain with Eq.~\eqref{eq:gaussian-PDF} an adjustment to the data that is nothing more than a mean. For example, if we want to estimate $\mathbb{G}m_k$, where $m_k$ is the $k$-th element of the parameter vector $\mathbf{m}$, from Eq.~\eqref{eq:gaussian-PDF} we find the following stationary point
\begin{equation}
    \begin{split}
        \frac{\partial\, p(m_k)}{\partial\, m_k} &= \frac{\partial}{\partial\, m_k}\frac{1}{\sqrt{2\pi}}\sum_{i=1}^{N}\exp{\left(-\frac{1}{2}\big( d_i - \mathbb{G}m_k\big]^2 \right)}\\
        &= 0\\
        &\Rightarrow \mathbb{G}m_k = \frac{1}{N}\sum_{i=1}^N d_i
    \end{split}
    \label{eq:gaussian-estimative}
\end{equation}

Therefore, if the observed data is contaminated with outliers, what we find using the conventional approach is the measurement of a mean, which in turn is strongly influenced by outliers and even more by a large amount of outliers. 

For an objective function to admit a minimum, it must satisfy the condition
\begin{equation}
    \sum_{i=1}^{N} c_i \mathcal{I}(x_i) = 0, \quad \textrm{with } \mathcal{I}(x) := \frac{\partial\, \phi (x)}{\partial m_k}
    \label{eq:influence-fuction}
\end{equation}

In Eq.~\eqref{eq:influence-fuction} $c_i$ are arbitrary constants and $\mathcal{I}$ is the so-called influence function \cite{Lima2021}. The influence function Eq.~\eqref{eq:influence-fuction} informs us about the sensitivity of the objective function to outliers. In this sense, if $\mathcal{I} \rightarrow \infty$ for a certain observed data $d^{obs}_i \rightarrow \infty$, the objective function is sensitive to outliers and, therefore, it is not a robust estimator. A robust estimate, however, will indicate $\mathcal{I} \rightarrow 0$ for $d^{obs}_i \rightarrow \infty$. What we have in the conventional approach, however, is that the conventional objective function, Eq.~\eqref{eq:classical-misfit}, is linearly influenced by outliers, as in the following equation
\begin{equation}
    \mathcal{I}(\mathbf{m}) = \sum_{i=1}^N x_i (\mathbf{m}) = -\sum_{i=1}^N \mathbb{G}^T \big( d^{obs}_i - \mathbb{G}\mathbf{m} \big).
    \label{eq:classic-influence}
\end{equation}

The conventional theory of inverse problems, based on Gaussian statistics, failures  with errors outside the Gaussian domain. In this sense, we look for alternatives to generalize Gaussian statistics in order to find more robust methods to deal with outliers.

\subsection{Rényi's Framework}
Based on  information theory, A. Rényi  \cite{Renyi1965informationTheory, renyi1961measures} introduced a  general information entropy as a one-parameter generalization of the BGS entropy. Rényi entropy ($\alpha$-entropy) functional is expressed by:
\begin{equation}
    \mathcal{S}_\alpha \big( p \big) = \frac{1}{1 - \alpha} \ln{ \left( \sum_{i=1}^N p^\alpha (x_i) \right) }, \quad \alpha \ge 0,
    \label{eq:renyi-entropy}
\end{equation}
where $\alpha$ is the entropic index. Furthermore, the entropy in Eq.~\eqref{eq:renyi-entropy} shares many of the properties of the BGS entropy Eq.~\eqref{eq:shannon-entropy}, such as: it is non-negative, it is additive, and it has an extreme for the case of equiprobability \cite{Renyi1965informationTheory}. The fundamental difference between these two entropies resides in the non-conservation of the concavity, which is associated to the choice of index $\alpha$. Furthermore, Rényi's entropy recovers Shannon's entropy at the limit $\alpha \rightarrow 1$. Applications of Rényi entropy can be found in several fields \cite{Wang2018,SnchezMoreno2010, Dong2016}.

\begin{figure}[!htb]
    \includegraphics[width=.9\columnwidth]{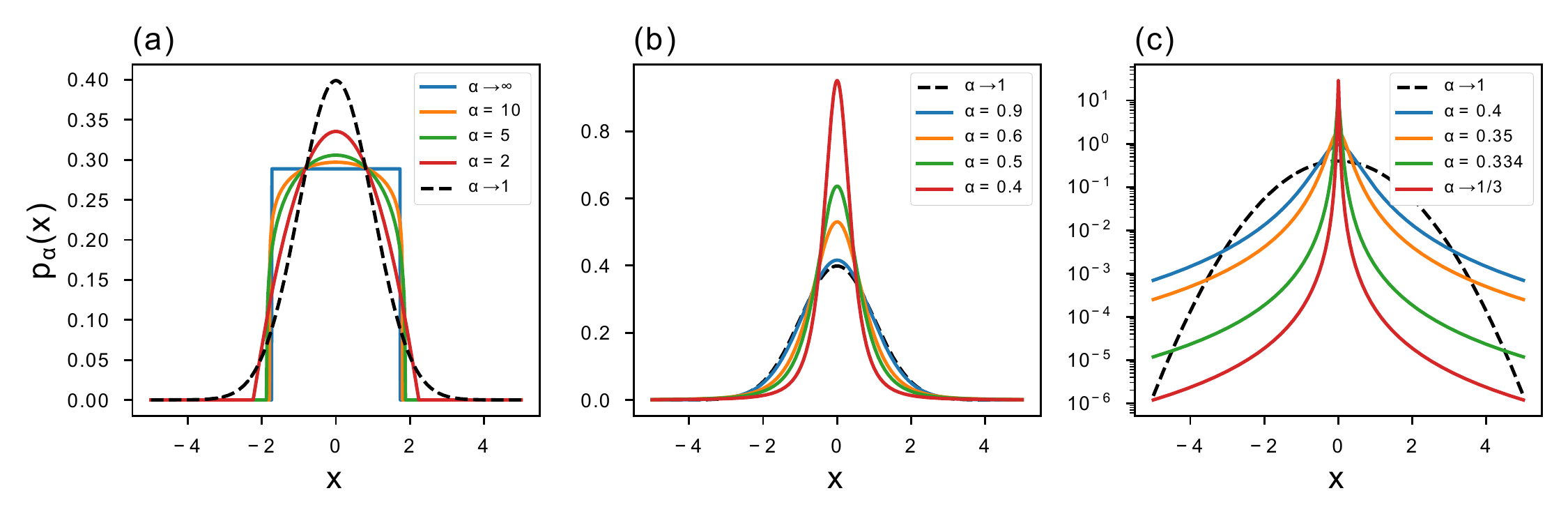}
    \caption{$\alpha$-Gaussian probability distribution. The black dashed line represents the conventional curves.}
    \label{fig:renyi-PDF}
\end{figure}

Taking into account the constraints in Eqs.~\eqref{eq:vinculos_norm} and \eqref{eq:vinculos_var}, $\alpha$-entropy is maximized by the $\alpha$-generalized  Gaussian distribution, $\alpha$-Gaussian, which is expressed in the form \cite{Costa2003, JOHNSON2007, Tanaka2019}: 
\begin{equation}
    p_\alpha (x) = A_\alpha \left( 1 - \frac{\alpha - 1}{3\alpha - 1} x^2\right)_{+}^\frac{1}{\alpha - 1}
    \label{eq:renyi-PDF}
\end{equation}
where $[x]_+ = 0$ if $x<0$ and $[x]_+ = x$. In addition,  $A_\alpha$ is the normalizing constant:
\begin{equation}
    A_\alpha =%
%    \begin{cases}
%    \sqrt{\frac{1 - \alpha}{[3\alpha - 1]\pi}}{\Gamma\left(\frac{1}{1 - \alpha}\right)}/{\Gamma\left(\frac{1+\alpha}{2[1 - \alpha]}\right)} \qquad \textrm{for} \quad 1/3 < \alpha < 1\\
%    \sqrt{\frac{1 - \alpha}{[3\alpha - 1]\pi}}{\Gamma\left(\frac{3\alpha - 1}{2[1 - \alpha]}\right)}/{\Gamma\left(\frac{\alpha}{\alpha-1}\right)} \qquad \textrm{for} \quad \alpha > 1
%    \end{cases}
    \begin{cases}
        \sqrt{\frac{1 - \alpha}{[3\alpha - 1]\pi}}{\Gamma\left(\frac{1}{1 - \alpha}\right)}/{\Gamma\left(\frac{1+\alpha}{2[1 - \alpha]}\right)}, \,\, \frac{1}{3} < \alpha < 1\\
        \sqrt{\frac{1 - \alpha}{[3\alpha - 1]\pi}}{\Gamma\left(\frac{3\alpha - 1}{2[1 - \alpha]}\right)}/{\Gamma\left(\frac{\alpha}{\alpha-1}\right)}, \,\, \alpha > 1
    \end{cases}
    \label{eq:cn-renyi}
\end{equation}
with $\Gamma (\cdot)$ representing the gamma function. At the limit $\alpha \rightarrow 1$, the ordinary Gaussian probability distribution Eq.~\eqref{eq:gaussian-PDF} is recovered. Figure~\ref{fig:renyi-PDF} shows some curves of the $\alpha$-Gaussian probability distribution. In particular, we note that at the limit $\alpha \rightarrow 1/3$  the probability distribution approaches a strongly peaked function.

Following the same path discussed of the previous Section, that is, using the maximum likelihood method, we find a generalized objective function \cite{sergio2021extensive}:
\begin{equation}
    \begin{split}
        \phi_\alpha (\mathbf{m}) &= \frac{1}{1-\alpha} \sum_{i=1}^{N} \ln{\left( 1 - \frac{\alpha - 1}{3\alpha - 1} x^2_i(\mathbf{m}) \right)_+}\\
        &= \| x_i \|_\alpha^2.
    \end{split}
    \label{eq:misfit-renyi}
\end{equation}

This function will be called $\alpha$-objective function and at limit $\alpha \rightarrow 1$ the conventional objective function, Eq.~\eqref{eq:classical-misfit}, is recovered. 
To investigate the behaviour of the  $\alpha$-objective function regarding outliers  we compute the influence function, as defined in Eq.~\eqref{eq:influence-fuction}, named  $\alpha$-influence function:
\begin{equation}
    \mathcal{I}_{\alpha}( \mathbf{m} ) =  \sum_{i=1}^N \frac{2x_i(\mathbf{m})}{\left( 3\alpha - 1 - (\alpha - 1)x_i^2(\mathbf{m}) \right)_+}
    \label{eq:renyi-influence}
\end{equation}

\begin{figure*}[!htb]
    \resizebox{\textwidth}{!}{\includegraphics{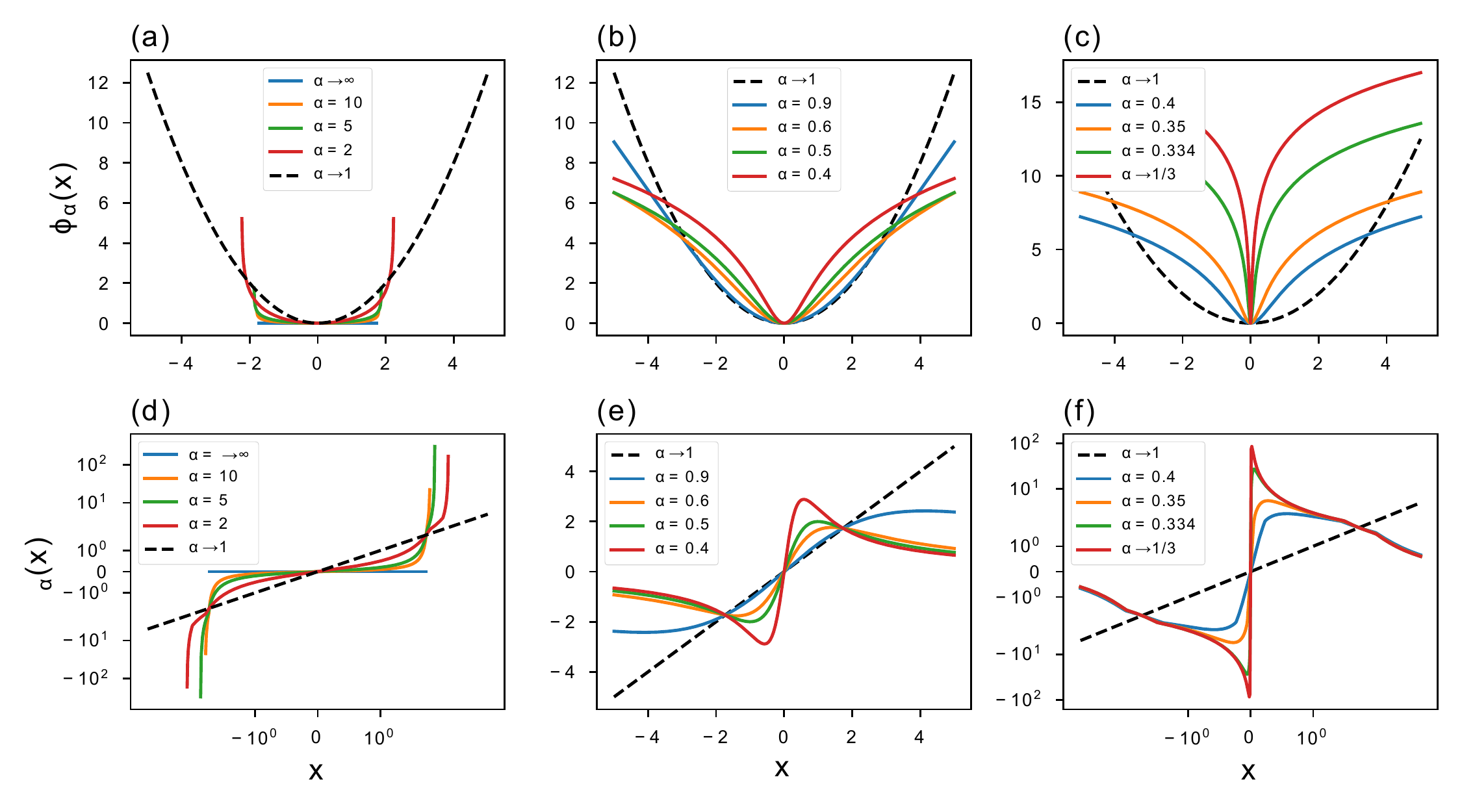}}
    \caption{(a)-(c) objective functions and (d)-(f) influence functions generalized based on Rényi statistic. The black dashed line represents the conventional curves.}
    \label{fig:renyi-misfit}
\end{figure*}

A couple of illustrative curves of the influence function are shown in Fig.~\ref{fig:renyi-misfit}, we draw our attention to the limit  case $\alpha \rightarrow 1/3$. In this region the influence of the outliers is minimized: $\mathcal{I}_\alpha (x_i \rightarrow \pm\infty) = 0$,  in contrast, the conventional objective function Eq.~\eqref{eq:classical-misfit} is strongly influenced by outliers since  $\mathcal{I}(x_i \rightarrow \pm\infty) = \pm\infty$.

\subsection{Tsallis's Framework}
Based on multifractals quantities and long-range interactions, C. Tsallis postulates an alternative  form for the entropy to generalize the standard statistical mechanics \cite{Tsallis1988}. Since then, a wide variety of applications have been performed based on Tsallis $q$-statistics \cite{PICOLIJR2009,DASILVA2021125539,Erick_2021_q_Pareto_EPBJ,DASILVA_CORSO_2021_Marsquakes}. The Tsallis approach is based on the $q$-entropy, defined as follows:
\begin{equation}
    \mathcal{S}_q \big( p \big) = \frac{1}{q - 1}\left( 1 - \sum_{i=1}^N p^q(x_i) \right),
    \label{eq:entropy-tsallis}
\end{equation}
where $q \in \mathbb{R}$ is the entropic index (also known as nonextensive parameter). The choice of the entropic index $q$ assigns new properties to the entropy functional, and in the limit case $q \rightarrow 1$ it recovers  the conventional BGS entropy. 

By considering the maximum entropy principle for $q$-entropy, a $q$-generalization of Gauss' law of error was formulated in Ref.~\cite{Suyari2005} assuming that the errors $x$ follow an optimal probability distribution. In this regard, the optimal probability function is computed by maximizing the $q$-entropy constrained to the normalization condition, Eq.~\eqref{eq:vinculos_norm}, and the $q$-variance \cite{qlogPlastino, qExpectation} given by:
\begin{equation}
    \langle x\rangle_q^2 = \sum_{i=1}^N x_i^2P_q(x_i), \,\, \textrm{with} \, P_q (x_j) = \frac{p^q (x_j)}{\sum_{i=1}^N p^q (x_i)},
\end{equation}
in which $P_q$ is the escort probability function \cite{Schlogl1993thermodynamics, abe2000remark}. The probability distribution resulting from the aforementioned optimization problem is known by the  $q$-Gaussian distribution:
\begin{equation}
    p_q(x) = A_q \left( 1 + \frac{q-1}{3-q}x^2 \right)_+ ^{\frac{1}{1-q}}
    \label{eq:tsallis-PDF}
\end{equation}
where the normalization constant is given by \cite{prato1999nonextensive}: 
\begin{equation}
    A_q =%
    \begin{cases}
%    \sqrt{\frac{1-q}{[3-q]\pi}}{ \Gamma\left( \frac{5-3q}{2[1 - q]} \right) }/{\Gamma\left( \frac{2-q}{1-q} \right)} \qquad \textrm{for} \quad -\infty < q < 1\\
%     \sqrt{\frac{q-1}{[3-q]\pi}}{ \Gamma\left( \frac{1}{q-1} \right) }/{\Gamma\left( \frac{3-q}{2[q-1]} \right)} \qquad \textrm{for} \quad 1 < q < 3\\
        \sqrt{\frac{1-q}{[3-q]\pi}}{ \Gamma\left( \frac{5-3q}{2[1 - q]} \right) }/{\Gamma\left( \frac{2-q}{1-q} \right)},  \,\, -\infty < q < 1\\
        \sqrt{\frac{q-1}{[3-q]\pi}}{ \Gamma\left( \frac{1}{q-1} \right) }/{\Gamma\left( \frac{3-q}{2[q-1]} \right)}, \,\, 1 < q < 3\\
    \end{cases}
    \label{eq:tsallis-cn}
\end{equation}

\begin{figure}
    \resizebox{.9\columnwidth}{!}{\includegraphics{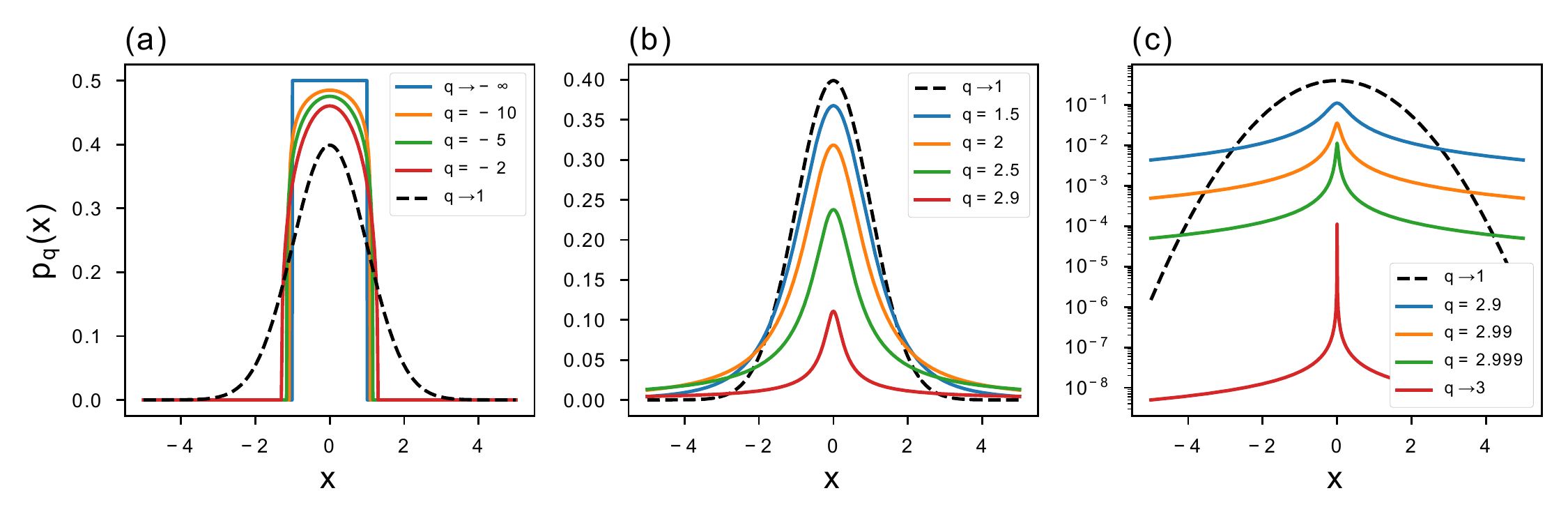}}
    \caption{$q$-Gaussian probability distribution. The black dashed line represents the conventional curves.}
    \label{fig:tsallis-PDF}
\end{figure}

A comparison between the conventional Eq.~\eqref{eq:shannon-entropy} and Tsallis approach Eq.~\eqref{eq:entropy-tsallis} reveals that   most probable events gain greater weight in the entropy calculation for the case in which $ q \ne 1 $. In this sense, the usual average is replaced by an average that depends on the choice of the index and so the higher the value of this index \cite{hasegawa2009properties}, the most likely events receive higher weights. Figure.~\ref{fig:tsallis-PDF} show illustrative curves of the $q$-Gaussian probability distribution. It is important to note that at the limit $q \rightarrow 3$ we have a behaviour that reminds us of the Rényi distribution in $\alpha \rightarrow 1/3$: at this limit both distributions display a peaked behaviour.

Applying the probabilistic maximum-likelihood method in the $q$-Gaussian distribution, we have the following objective function \cite{daSilva2020robust}
\begin{equation}
    \begin{split}
        \phi_q (\mathbf{m}) &= \frac{1}{q-1} \sum_{i=1}^N \ln{\left( 1 + \frac{q-1}{3-q}x_i^2(\mathbf{m}) \right)_+}\\
        &= \| x_i \|_q^2
    \end{split}
    \label{eq:tsallis-misfit}
\end{equation}

In order to check the influence of outliers for our objective function, we calculate the influence function \cite{Lima2021}:
\begin{equation}
    \mathcal{I}_{q}(\mathbf{m}) = \sum_{i=1}^N \frac{2x_i(\mathbf{m})}{\left(3-q+(q-1)x_i^2(\mathbf{m})\right)_+}
    \label{eq:tsallis-influence}
\end{equation}

\begin{figure*}[!htb]
    \resizebox{\textwidth}{!}{\includegraphics{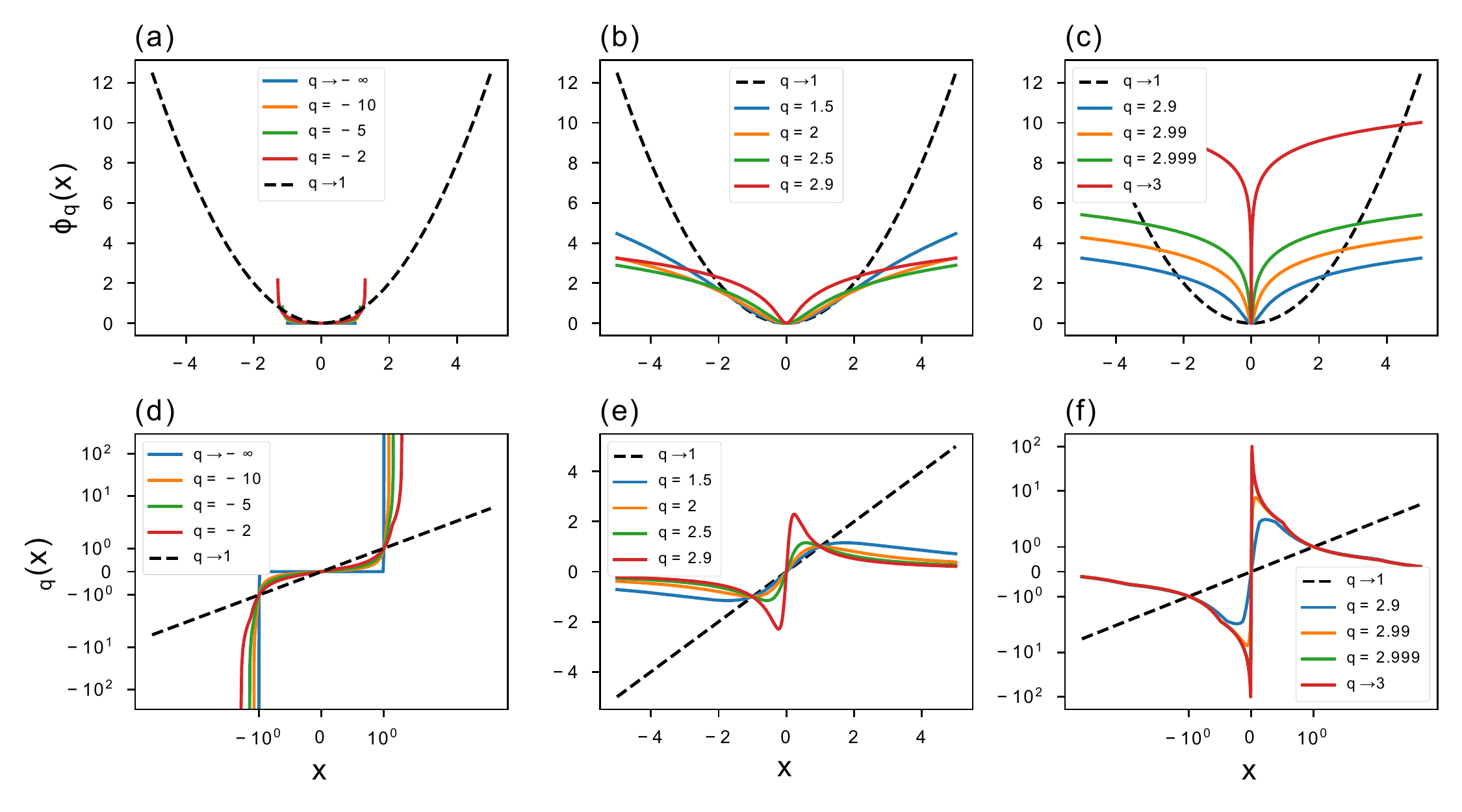}}
    \caption{(a)-(c) objective functions and (d)-(f) generalized influence functions based on Tsallis statistic. The black dashed line represents the conventional curves.}
    \label{fig:tsallis-misfit}
\end{figure*}

Equation~\eqref{eq:tsallis-influence} reveals that the Tsallis framework also shows a robust objective function that is resistant to outliers. Figure~\ref{fig:tsallis-misfit} displays a couple of  influence function curves. At the limit $x_i \rightarrow \pm\infty$, the choice of index $q < 1$ implies $\left(3-q+(q-1)x_i^2\right)_+ ^{-1} \rightarrow \infty$ because the sum inside the brackets will always result in a negative quantity: $\mathcal{I}_{q<1}(\pm \infty) \rightarrow \infty$. On the other hand, for $q>1$ the sum inside the brackets turns into a large positive number, leading to $\mathcal{I}_{q>1}(\pm\infty) \rightarrow 0$.

\subsection{Kaniadakis's Framework}
G. Kaniadakis proposed a new way to calculate the  entropy based on the principle of Kinetic Interaction \cite{kaniadakis2001,  PhysRevEkaniadakis2002}. This new $\kappa$-entropy that    generalizes the  BGS statistics is given by: 
\begin{equation}
    \mathcal{S}_\kappa \big( p\big) = -\frac{1}{2\kappa}\sum_{i=1}^N{ \Big( p^{1+\kappa}(x_i) - p^{1 - \kappa}(x_i)\Big)}
    \label{eq:kaniadakis-entropy}
\end{equation}
Kaniadakis statistics has been applied in different contexts \cite{Kaniadakis2020_epidemologia,DASILVA2021110622,Kaniadakis2021_powerlaw}. The  conventional entropy (BGS) is recovered in the limit of the entropic index $\kappa \rightarrow 0$. Kaniadakis' framework not only includes conventional BGS statistics, but it is related to other statistics, such as the famous quantum statistics of Fermi-Dirac and Bose-Einstein, as well as the Tsallis \cite{kaniadakis2001, Santos2011}.

Based on the $\kappa$-Gaussian statistic, the reference \cite{sergio2021kaniadakis} presents an error law that can be applied to a variety of problems and that, because it has a heavy tails distribution, it is able to satisfactorily work with outliers. The $\kappa$-Gaussian distribution is given by
\begin{equation}
    p_\kappa (x) = \frac{1}{A_\kappa} \Bigg(\sqrt{1 + \kappa^2\beta_\kappa^2 x^4} - \kappa \beta_\kappa x^2 \Bigg)^{1/\kappa}
    \label{eq:kaniadakis-PDF}
\end{equation}
where $A_\kappa$ is the normalization constant given by
\begin{equation}
    A_\kappa = \Big(1 + \frac{\mid\kappa\mid}{2} \Big)\sqrt{\frac{2\mid\kappa\mid\beta_\kappa}{\pi}}\frac{\Gamma\left( 1/\mid2\kappa \mid + 1/4\right)}{\Gamma\left( 1/\mid2\kappa \mid - 1/4\right)}
\end{equation}
and $\beta_\kappa > 0$ depends on the $\kappa$ index and is given by:
\begin{equation}
\begin{split}
    \beta_\kappa &= \frac{1 + \mid\kappa\mid/2}{\mid 2\kappa\mid(2 + 3\mid\kappa\mid)}\\
    &\times\frac{\Gamma(1/\mid2\kappa\mid - 3/4)}{\Gamma(1/\mid2\kappa\mid + 1/4)}%
    \frac{\Gamma(1/\mid2\kappa\mid + 3/4)}{\Gamma(1/\mid2\kappa\mid - 1/4)}    
\end{split}
%    \beta_\kappa = \frac{1 + \mid\kappa\mid/2}{\mid 2\kappa\mid(2 + 3\mid\kappa\mid)}%
%    \frac{\Gamma(1/\mid2\kappa\mid - 3/4)}{\Gamma(1/\mid2\kappa\mid + 1/4)}%
%    \frac{\Gamma(1/\mid2\kappa\mid + 3/4)}{\Gamma(1/\mid2\kappa\mid - 1/4)}
\end{equation}

\begin{figure}[!htb]
    \centering
    %\resizebox{\columnwidth}{!}{\includegraphics{fig_06.pdf}}
    \includegraphics[width=14cm]{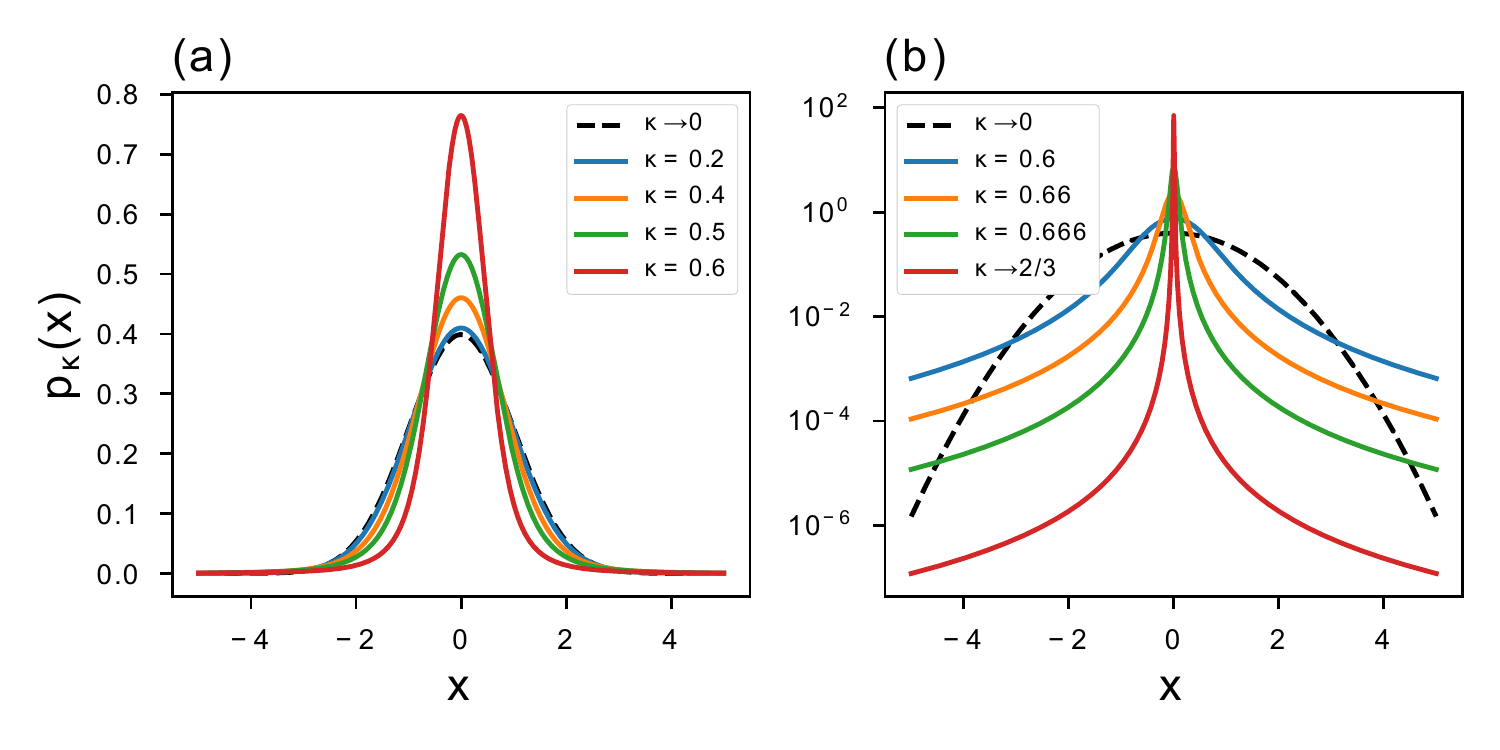}
    \caption{$\kappa$-Gaussian probability distribution. The black dashed line represents the conventional curves.}
    \label{fig:kaniadakis-PDF}
\end{figure}

Some curves of the $\kappa$-Gaussian distribution are shown in Fig.~\ref{fig:kaniadakis-PDF}. We notice that the distribution has heavy tails when choosing $\kappa \rightarrow 2/3$ and at this limit, as well as in the $\alpha$-Gaussian and $q$-Gaussian distributions, the distribution shows a peaked behaviour that resembles  the Dirac delta distribution. 

In this scenario, the inverse problem is therefore formulated as the problem of optimizing the $\kappa$-objective function that derives from the principle of maximum likelihood
\begin{equation}
    \begin{split}
        \phi_\kappa (\mathbf{m}) &= -\frac{1}{\kappa} \sum_{i=1}^N \ln{\Bigg( \sqrt{1 + \kappa^2 \beta_\kappa^2 x_i^4(\mathbf{m})}
        - \kappa \beta_\kappa x_i^2(\mathbf{m})\Bigg)}\\ 
        &= \| x_i \|_\kappa^2
    \end{split}
    \label{eq:misfit-kaniadakis}
\end{equation}
and the analysis of the robustness of this objective function can be performed by the $\kappa$-influence function, which is given by:
\begin{equation}
    \mathcal{I}_{\kappa}(\mathbf{m}) = \sum_{i=1}^N \frac{2\beta_\kappa x_i(\mathbf{m})}{\sqrt{ 1 + \kappa^2\beta_\kappa^2 x_i^4(\mathbf{m}) }}
    \label{eq:influence-kaniadakis}
\end{equation}

\begin{figure*}[!htb]
    \centering
    \includegraphics[width=.8\textwidth]{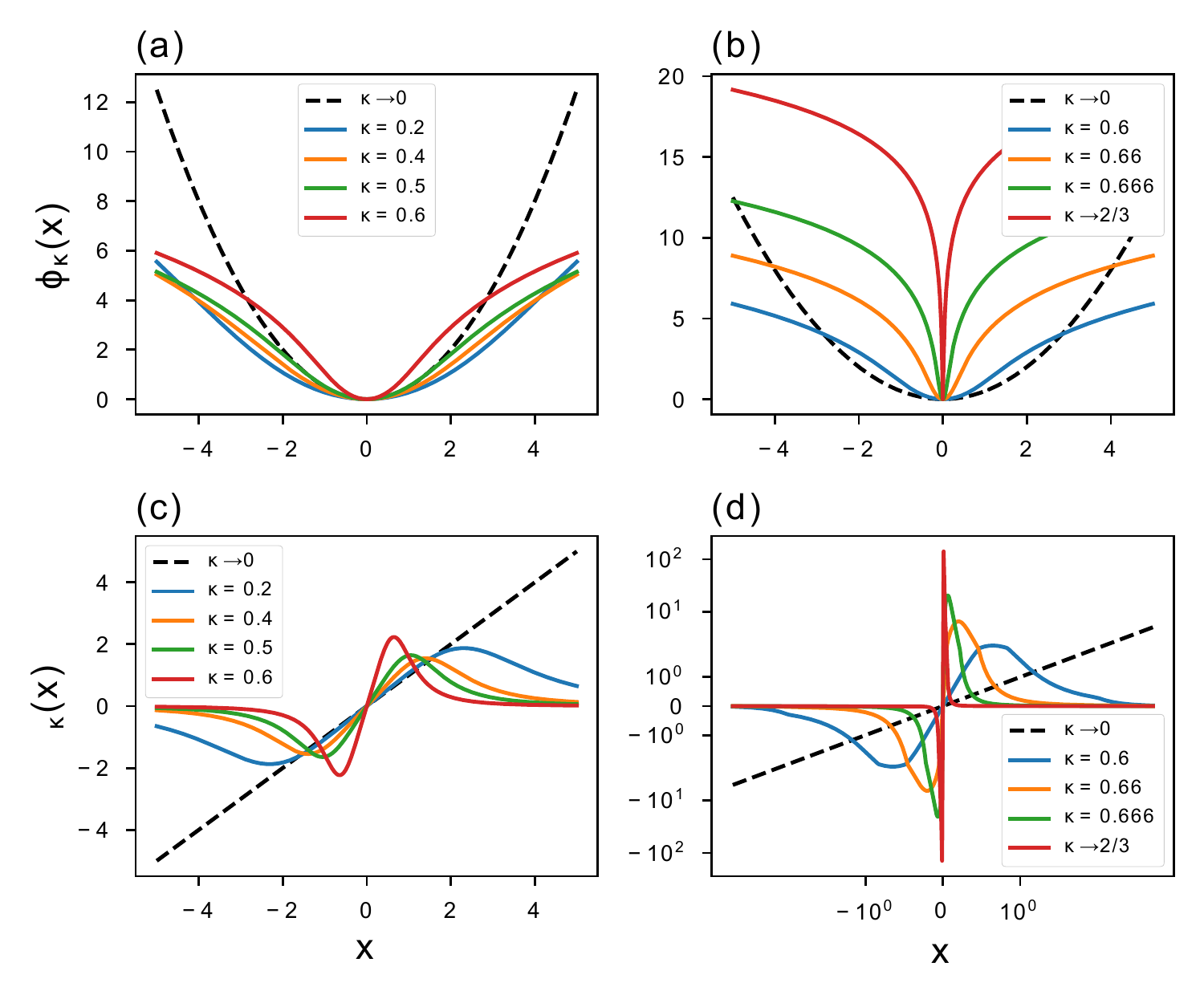}
    \caption{(a)-(b) objective functions and (c)-(d) generalized influence functions based on Kaniadakis statistic. The black dashed line represents the conventional curves.}
    \label{fig:kaniadakis-misfit}
\end{figure*}

The curves of the $\kappa$-objective and $\kappa$-influence functions are shown in Fig.~\ref{fig:kaniadakis-misfit}. In Fig.~\ref{fig:kaniadakis-misfit}(c)-(d) we notice that as we increase the value of index $\kappa$ the influence of distant values of $x=0$ decreases. In particular, at the limit $\kappa \rightarrow 2/3$ we observe the curve that indicates less influence for $x \rightarrow \pm \infty$ measurements. Looking at Eq.~\eqref{eq:influence-kaniadakis} we noticed that for $x_i \rightarrow \pm \infty$ we have $\mathcal{I}_\kappa (x_i \rightarrow \pm \infty) = 0$.

\subsection{Comparing the performance of objective functions}\label{sec:comparacao}
In this section, we present a simple numerical experiment in order to quantitatively analyse the robustness of objective functions based on generalized statistics. The experiment consists of estimating the coefficients $\mathbf{m} = \{m_1 , m_2\}$ of a linear polynomial, $d^{cal} = m_1 x + m_2$, from observed data $\mathbf{d}^{obs}$ contaminated by outliers. In this regard, we consider the independent variable $x \in \mathbb{R}^{50}$ within the range $[-1, 1]$ to generate $50$ numbers obeying a linear polynomial with coefficients $m_1 = 1$ and $m_2 = 2$. Then we contaminate the numbers generated with a Gaussian distribution with zero-mean and standard-deviation $\sigma^2 = 0.2$. In addition, the variable $d^{obs}$ is contaminated with outliers  in region $0.4 \le x < 0.9$, of $d^{obs}$, the outliers are given by $d^{obs}_i = 10f$ where $f$ is a Gaussian random variable.

Conventionally, this problem is treated by minimizing the square of the residuals based on  Gaussian statistics \cite{MENKE20121, weisberg2005applied}. However, as discussed in Eq.~\eqref{eq:gaussian-estimative}, the Gaussian estimate is not appropriate to solve problems with discrepant values. In this sense, we propose to estimate the coefficients $\mathbf{m}$ using the objective functions of Rényi (Eq.~\eqref{eq:misfit-renyi}), Tsallis (Eq.~\eqref{eq:tsallis-misfit}) and Kaniadakis (Eq.~\eqref{eq:misfit-kaniadakis}). The values of the entropic index were used between $1\le \alpha \le 0.3334$, $1 \le q \le 2.9999$ and $0 \le \kappa \le 0.6666$. Inside each interval, $200$  uniform spaced  values were taken.

\begin{figure*}[!htb]
    \resizebox{\textwidth}{!}{\includegraphics{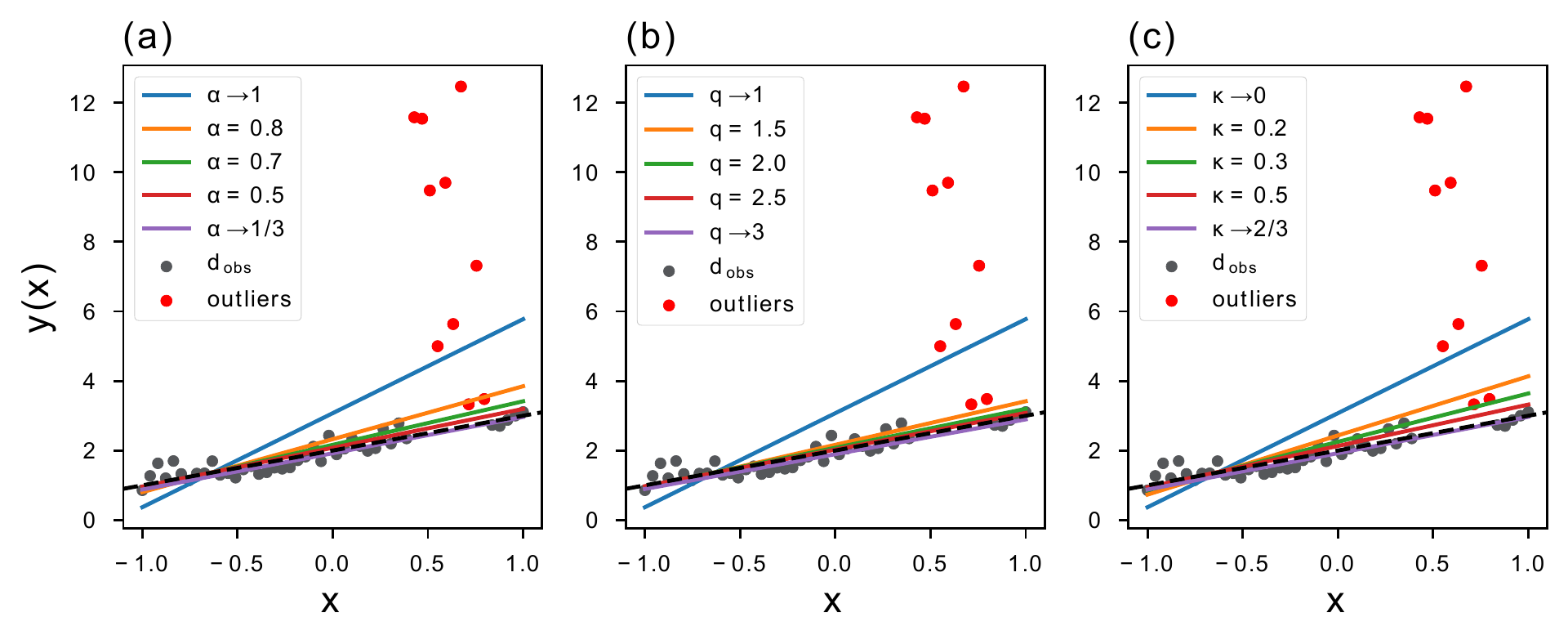}}
    \caption{Fit of the estimated lines using the objective functions of (a) Rényi, (b) Tsallis and (c) Kaniadakis. The dashed line indicates an ideal line, and the points highlighted in red indicate the inserted outliers.}
    \label{fig:expA_fitCurves}
\end{figure*}

\begin{figure*}[!htb]
    \resizebox{\textwidth}{!}{\includegraphics{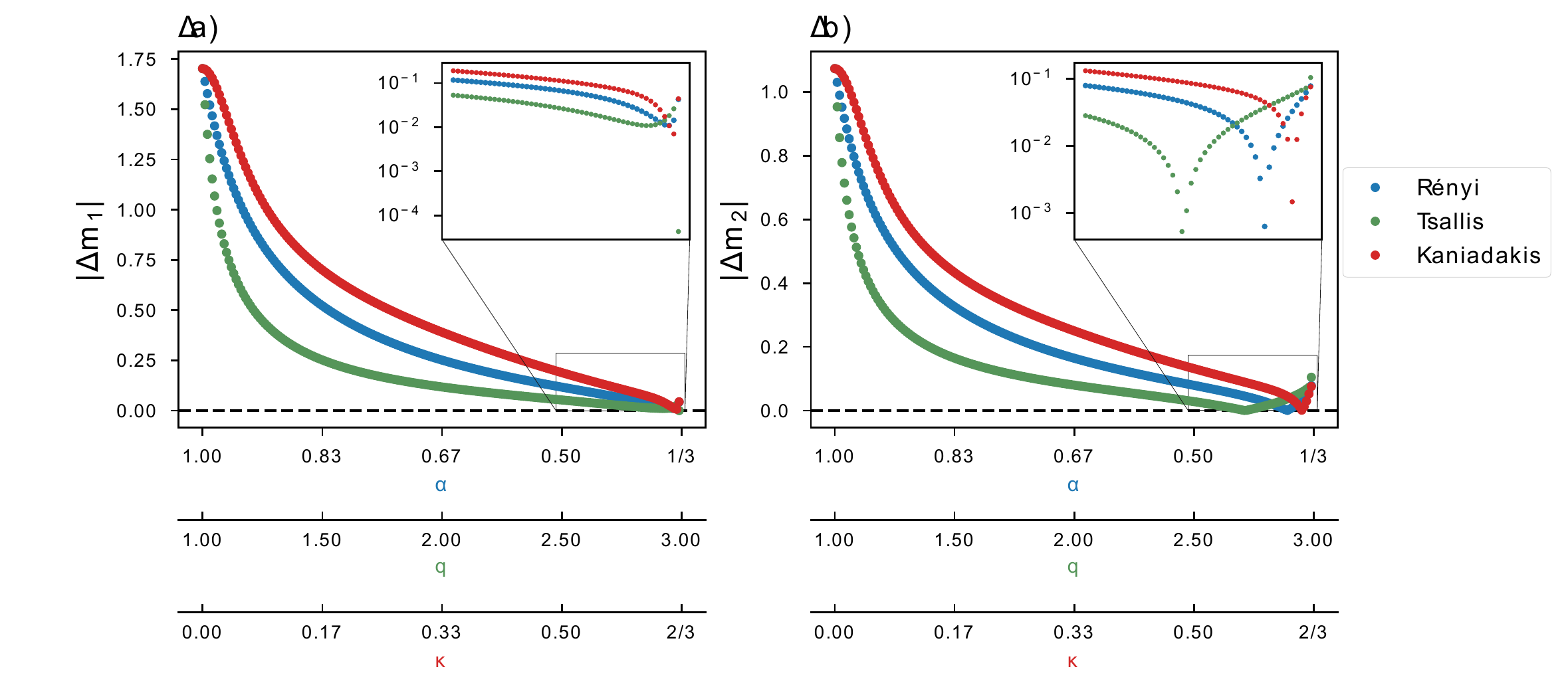}}
    \caption{Relation between estimated parameters and entropic indexes. The zoom in window in (a) and (b) emphasizes the regions $0.5 \le \alpha < 1/3$, $2.5 \le q < 3$ and $0.5 \le \kappa < 2/3$}
    \label{fig:expA_mEstimative}
\end{figure*}

The calculated models, $\mathbf{m}^{cal} = \{m_1^{cal},\,m_2^{cal}\}$, are compared with the ideal model $\mathbf{m} = \{1,\,2\}$  to find  for each objective function the best entropic index that achieve an optimal fitting. Figure~\ref{fig:expA_mEstimative} correlates the estimated values of the intercept, $\Delta m_1 = m_1^{cal} - 1$, and the slope, $\Delta m_2 = m_2^{cal} - 2$, with the employed entropic index. We can see that as we move away from the Gaussian limit, we find a better fit.

We compared the lines obtained with the estimated parameters, calculating the Mean Absolute Error, $MAE = \sum_i \mid d_i(\mathbf{m}^{cal}) - d_i(\mathbf{m})\mid/N$, between the calculated line, $d(\mathbf{m}^{cal})$, and the ideal line, $d(\mathbf{m})$. The obtained results are:  $MAE_\alpha = 0.0136$ with index $\alpha = 0.3635$; $MAE_q = 0.0134$ with $q = 2.7587$ and $MAE_\kappa = 0.0127$ with $\kappa = 0.6532$. The $MAE$ represents the average deviation between predicted and reference values, and the best result is achieved for $MAE$ close to zero. In contrast, the $MAE$ calculated with the conventional objective function was $1.2000$.

With this test, we observe that the results improve as we move away from the Gaussian limit. In particular, we notice that in Fig.~\ref{fig:expA_mEstimative} the green curve corresponding to the Tsallis result  shows the steepest decrease  after the Gaussian limit ($q = 1$), while the red curve (Kaniadakis) presents a curve that slowly decreases after the Gaussian limit.

An analysis of the influence functions, Eqs.~\eqref{eq:renyi-influence}, \eqref{eq:tsallis-influence} and \eqref{eq:influence-kaniadakis}, reveals that they are zero for $x_i \rightarrow \pm \infty$, regardless of the choice of entropic indexes (obviously, disregarding the conventional limit). In addition, we  notice that in the limits $\alpha \rightarrow 1/3, \, q \rightarrow 3$ and $\kappa \rightarrow 2/3$ the influence functions are merely a function that depends on the inverse of $x_i$. Thus, the three objective functions, Eqs.~\eqref{eq:misfit-renyi}, \eqref{eq:tsallis-misfit} and \eqref{eq:misfit-kaniadakis}, are resistant to outliers and have an entropic index limit in which they are equivalent.

\section{Numerical experiments} \label{sec:numericalExperiment}

In this section, we present numerical experiments to demonstrate the outlier-resistance of the data-inversion method based on generalized statistics by considering an important problem that comes from geophysics, which is an important process to obtain estimates of subsurface properties. In particular, we address a problem of seismic inversion known as Post-Stack Inversion (PSI) \cite{psi}. The goal of PSI is to estimate, from the observation of the seismic data, the acoustic impedance, which is a property of the rock defined as the product of the density of the rock and the speed of the acoustic wave in the subsurface \cite{sen2006seismic}.

The forward problem is formulated through the following relationship: $\mathbf{d}^{cal} = \mathbb{W D}\mathbf{m}$. Here,  $\mathbf{d}^{cal}$  represents the seismic data calculated by the parameters $\mathbf{m} = \ln{(\mathbf{Z})}$ which is used to estimate the acoustic impedance $\mathbf{Z}$. The operators $\mathbb{W}$ and $\mathbb{D}$ are described by \cite{wu2020seismic}: 
\begin{equation}
    \mathbb{W} = %
   \begin{bmatrix}%
        \omega_1	&		0		&	...	    &	0\\
        \vdots		&	\omega_1	&	\vdots  &	0\\	
        \omega_n	&	\vdots		&   \ddots  &	0\\
        0			& \omega_n		&  \vdots	&	\omega_1\\
        \vdots		&  \vdots		&  \ddots 	& \vdots \\
        0			& ...			&   ... 	& \omega_n
    \end{bmatrix},%
    \, %
    %\label{eq:matriz-W}
    \mathbb{D} = \frac{1}{2}%
    \begin{bmatrix}
        -1	    &	1       &	0    &	...	    &	0\\
        0   	&	-1	    &	1  	 & ...	    &   0\\
        \vdots	& \vdots	& \ddots & \ddots	&   0\\
        0	    & ...	    & ...	 & -1	    &   1
    \end{bmatrix}.
\end{equation}
Where $\mathbb{W}$ is  the wavelet operator, which computes the convolution between the seismic signal and   $\mathbb{D} \mathbf{m}$. Finally, $\mathbb{D}$ represents  the first order derivative operator. In addition, we can group these two matrices into  a single operator   $\mathbb{G} = \mathbb{W D}$. In this way, we compute the residuals between the observed data and the calculated
data  $\mathbf{x} = \mathbf{d}^{obs} - \mathbb{G} \mathbf{m}$.

\begin{figure*}
    \centering
    \includegraphics[width=.8\textwidth]{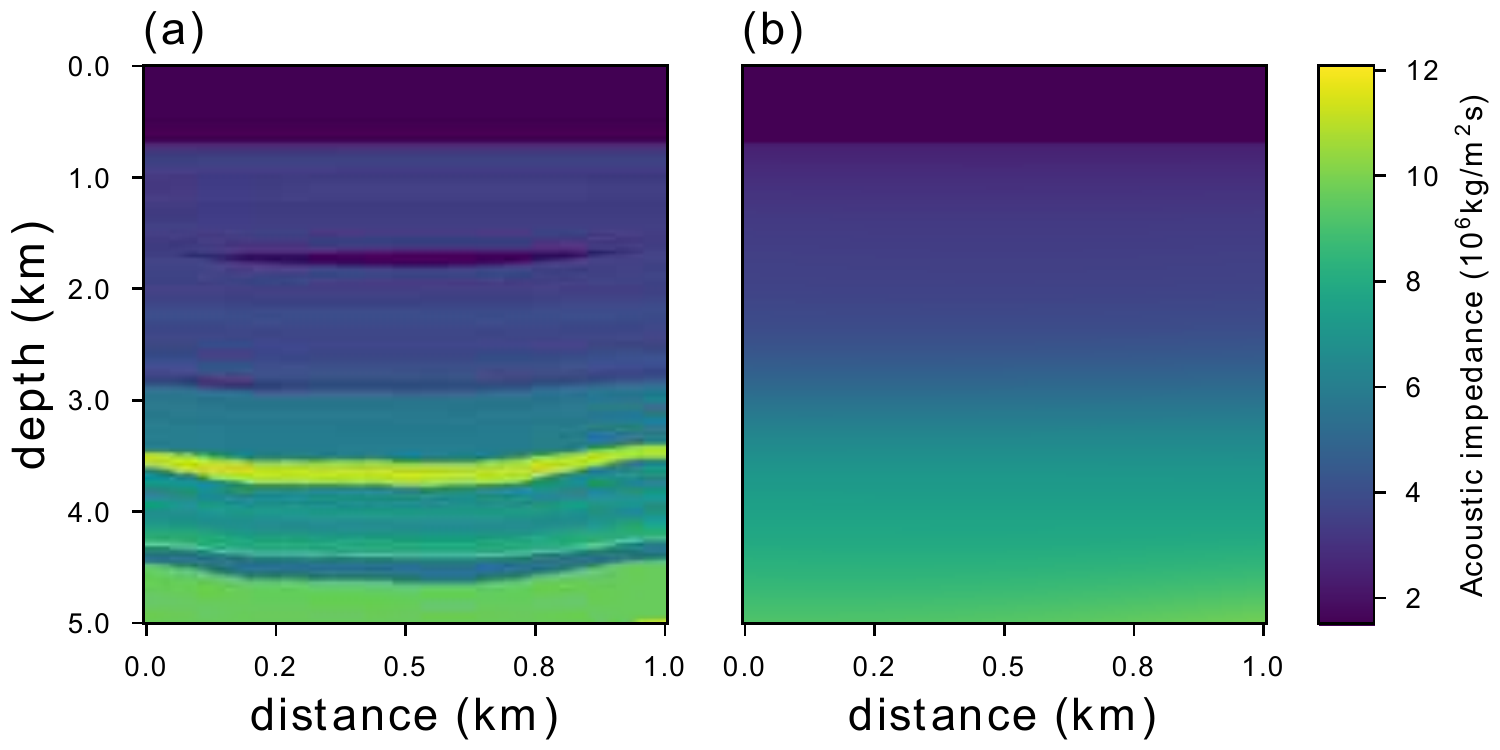}
    \caption{The geophysical model employed to illustrate the inversion methodology. In (a) the synthetic acoustic impedance model called Marmousi2. In (b) we show the  initial model employed in the inversion methodology.}
    \label{fig:AImodel}
\end{figure*}

To analyse the outlier-resistance of the objective functions presented in Section \ref{sec:generalizedStatisticsSeismic}, we consider a portion of the synthetic geological  Marmousi2  \cite{Versteeg_marmousi,MarmousiMartin} model as a benchmark (true model). In particular, we take into account the acoustic impedance model that consists of $5 \, km$ of depth and $1 \, km$ of distance as depicted in Fig.~\ref{fig:AImodel}. The seismic source used was a Ricker wavelet \cite{rickerSource, rickerSource2} with the peak frequency $\nu_p = 55 \, Hz$ (the most energetic frequency).

\begin{figure*}[!htb]
    \resizebox{\textwidth}{!}{\includegraphics{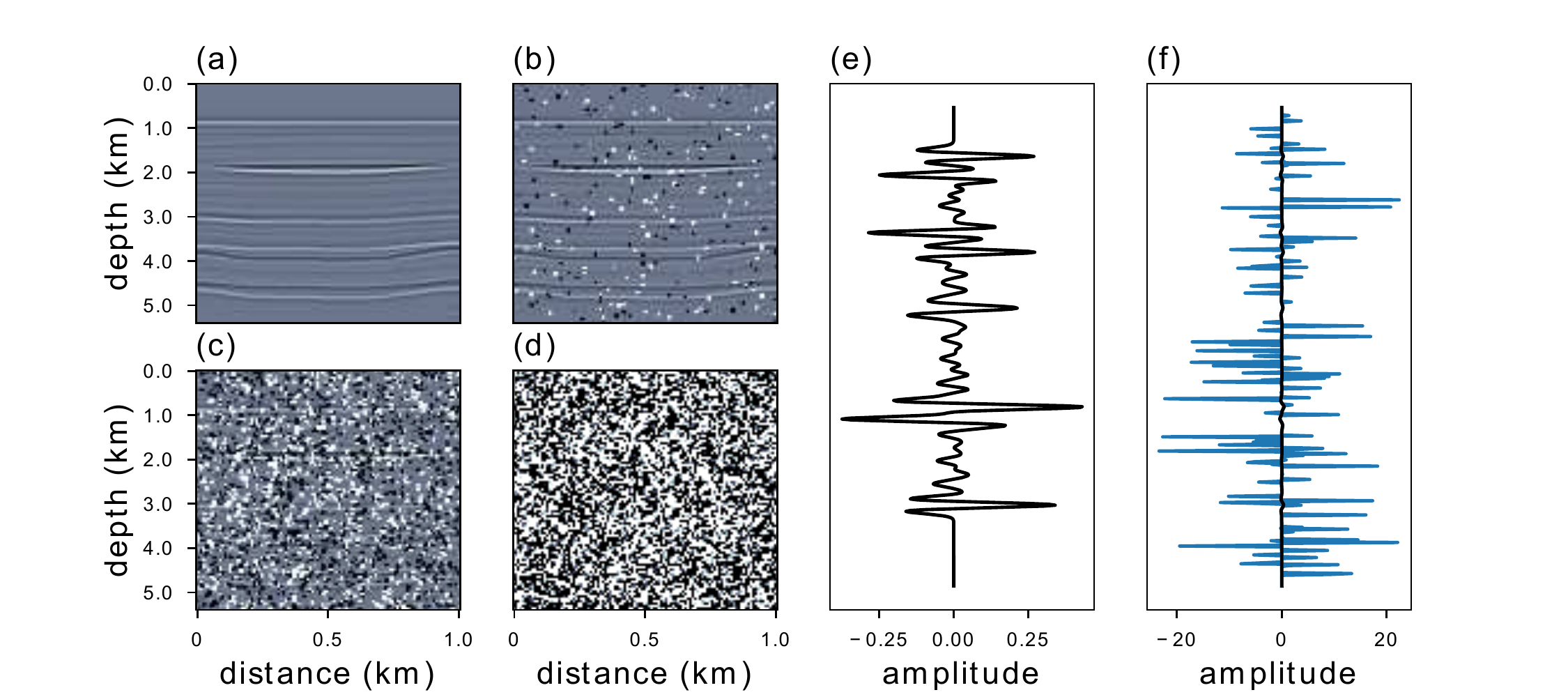}}
    \caption{The noiseless seismic data in  (a). The same data contaminated with white-noise (signal-to-noise ratio $SNR = 80$) and spike noise with (b) $0.5\%$, (c) $5\%$, and (d) $80\%$. The black line in panels  (e) represents a single seismic trace from the middle of panel (a). The same trace contaminated by noise is represented in (f) spikes ($25\%$.)}
    \label{fig:dobs}
\end{figure*}

We test the robustness of the generalized objective functions using seismic data contaminated with white Gaussian noise with low intensity (taking a signal-to-noise ratio equal to $80\,dB$ in all scenarios) and spikes (outliers) with different intensities, as shown in Fig.~\ref{fig:dobs}. The spikes were added by randomly choosing positions in the seismic data and adding peaks with intensities between $5f$ and $15f$ times the original amplitude, where $f$ is a Gaussian random variable. We considered $161$ noise scenarios, where the difference is in the percentages of samples contaminated by the outliers. In this regard, the number of samples  were chosen from $0 \%$ to $80\%$ of the data samples, with steps of $0.5\%$. 

For each spiky-noise scenario, we carried out data-inversions employing the $\alpha$-, $q$- and $\kappa$-objective functions, Eqs.~\eqref{eq:misfit-renyi}, \eqref{eq:tsallis-misfit} and \eqref{eq:misfit-kaniadakis}, respectively with $1/3 < \alpha \le 1$, $1 \le q < 3$ and $0 \le \kappa \le 2/3$ using 200 values for each of these parameters with uniform spacing between intervals. In total, for each objective function, we get $32.000$ results. To minimize the objective functions, we employ the conjugate gradient method \cite{stiefel1952methodsCG, scales1997geoinversetheory}, defining a maximum of 10 iterations and a tolerance error $\epsilon = 10^{-12}$.

We consider the  Pearson's correlation coefficient \cite{evans1996straightforward} as the statistical metric to compare the PSI results, which is defined as: 
\begin{equation}
    R = \frac{\sum_{n=1}^N\Delta^{true}_n\Delta^{rec}_n}{\sqrt{\sum_{n=1}^N\big(\Delta^{true}_n\big) ^2} \sqrt{\sum_{n=1}^N\big(\Delta^{rec}_n\big)^2} }
\end{equation}
where $\Delta^{true}_n = Z^{true}_n - \mu^{true}$ and $\Delta^{rec}_n = Z^{rec}_n - \mu^{rec}$ is the difference between the true and recovered acoustic impedance models, $Z$, and their respective averages, $\mu$. %
%\begin{equation}
%    R = \frac{\sum_{n=1}^N\big( Z_n^{true} - \mu_{true}\big)\big( Z_n^{rec} - \mu_{rec}\big)}{\sqrt{\sum_{n=1}^N\big( Z_n^{true} - \mu_{true}\big)^2} \sqrt{\sum_{n=1}^N\big( Z_n^{rec} - \mu_{rec}\big)^2} }
%\end{equation}
%where $Z^{true}$ and $Z^{rec}$ are the true and recovered acoustic impedance models and $\mu_{Z_{true}},\, \mu_{Z_{rec}} $ their respective averages.
The coefficient R assume values between $-1$ and $+1$. The case of R close to zero implies in absence of correlation. The correlation being   strong when it approaches one. 

\begin{figure*}[!htb]
    \resizebox{\textwidth}{!}{\includegraphics{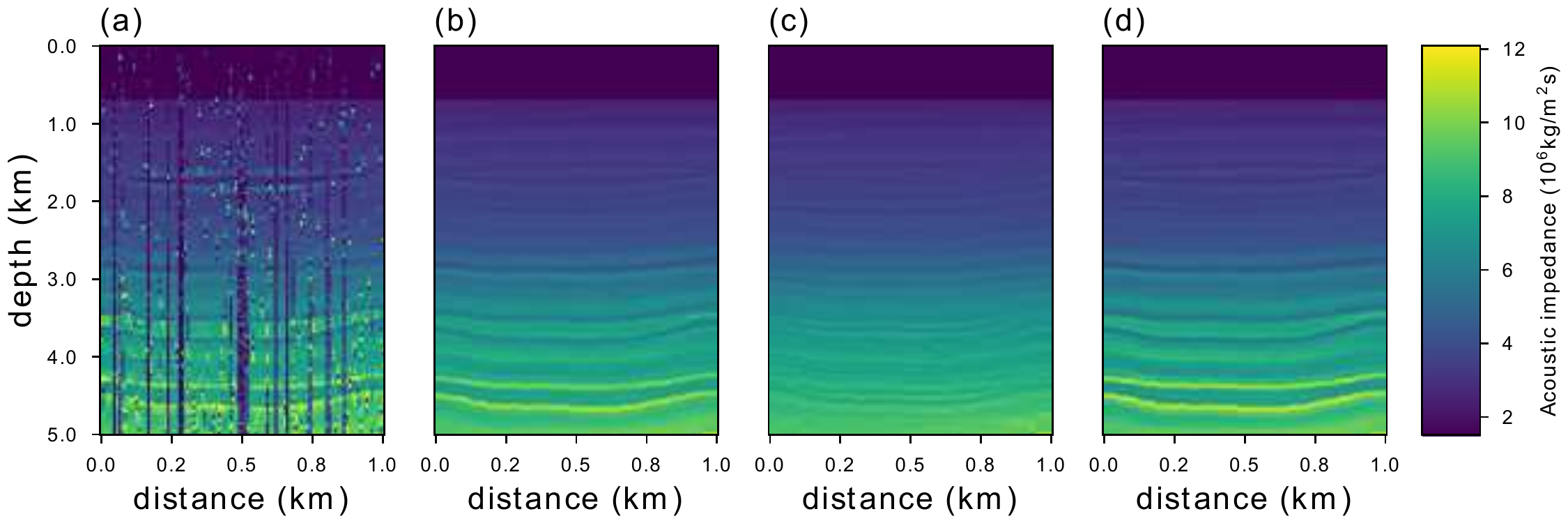}}
    \caption{Acoustic impedance model recovered for an observed data contaminated with white noise (SNR = 80) and spike noise ($0.5\%$) using objective function (a) conventional (b) Rényi with $\alpha = 0.3334$; (c) Tsallis with $q = 2.9999$; and (d) Kaniadakis with $\kappa = 0.6666$}
    \label{fig:AImodeled_spikePorcentagem_0.5}
\end{figure*}

\begin{figure*}[!htb]
    \resizebox{\textwidth}{!}{\includegraphics{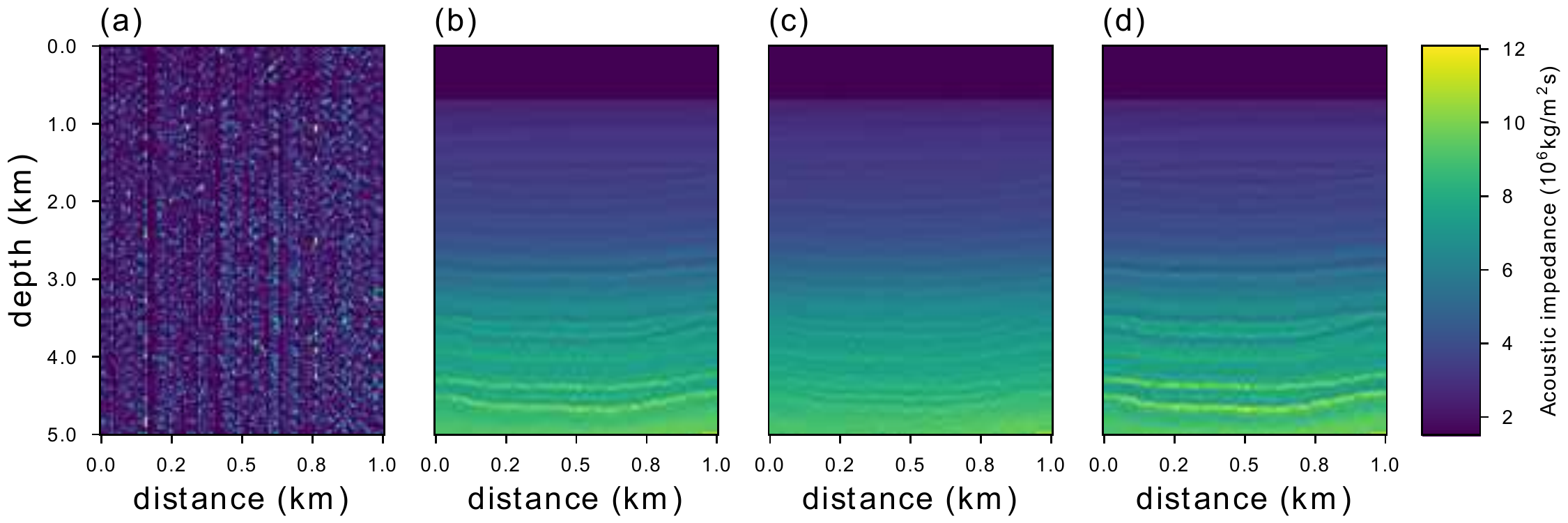}}
    \caption{Acoustic impedance model recovered for an observed data contaminated with white noise (SNR = 80) and spike noise ($80\%$) using objective function (a) conventional (b) Rényi with $\alpha = 0.3334$; (c) Tsallis with $q = 2.9999$; and (d) Kaniadakis with $\kappa = 0.6666$}
    \label{fig:AImodeled_spikePorcentagem_80}
\end{figure*}

Figures~\ref{fig:AImodeled_spikePorcentagem_0.5} and \ref{fig:AImodeled_spikePorcentagem_80} show the recovered acoustic impedance for   conventional and generalized objective functions. We notice that the PSI results obtained with the conventional objective function are severely affected by the presence of outliers. On the other hand, at the limits $\alpha \rightarrow 1/3$, $q \rightarrow 3$ and $\kappa \rightarrow 2/3$ the influence of spikes are minimized, and good estimates are obtained.

\begin{figure*}[!htb]
    \resizebox{\textwidth}{!}{\includegraphics{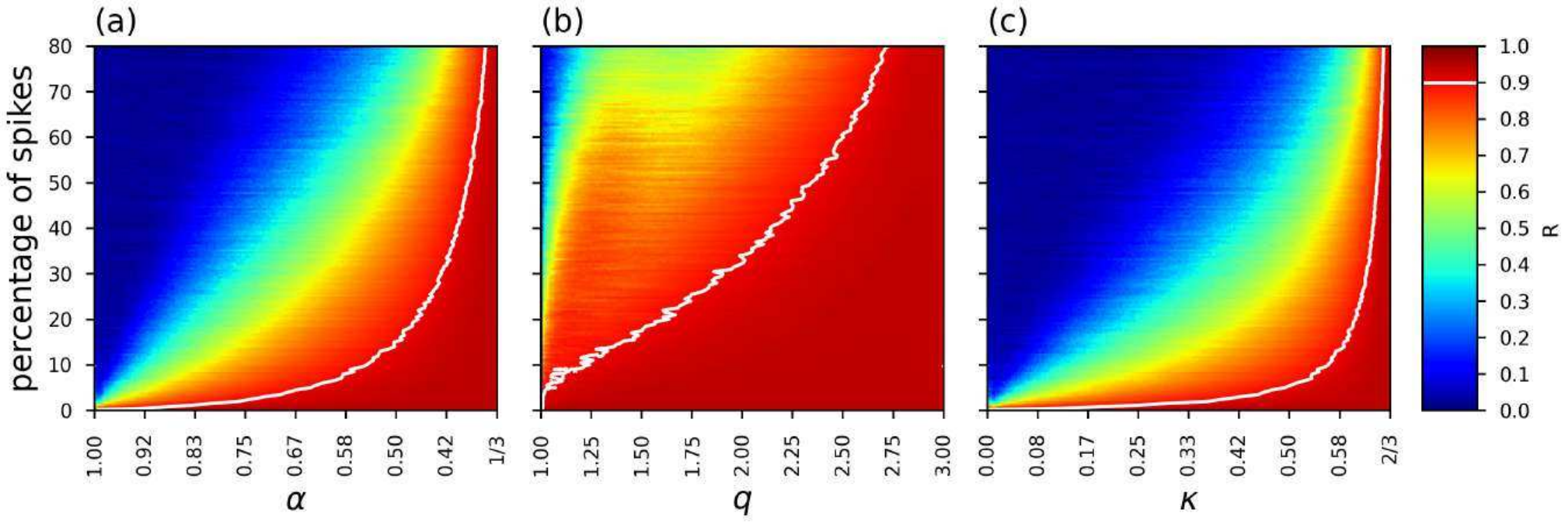}}
    \caption{Heatmap representing the correlation between synthetic and recovered impedance models for the objective functions: (a) Rényi, (b) Tsallis and (c) Kaniadakis. The white markings indicate points such that $R = 0.9$ (strong correlation).}
    \label{fig:heatmap_results}
\end{figure*}

We summarize the PSI results for all numerical simulations performed in the present work in Fig.~\ref{fig:heatmap_results}. In this figure, we remark that the objective functions present satisfactory results in $\alpha \rightarrow 1/3$, $q \rightarrow 3$ and $\kappa \rightarrow 2/3$ limit cases, as predicted by the numerical experiment presented in Section~\ref{sec:comparacao}. Indeed, in these limit cases, the generalized objective functions are robust tools capable of mitigate the influence of outliers, which  leads to good results even for high spike contamination. In addition, it should be noted that the PSI results related with the reddish regions of the heatmap in Fig.~\ref{fig:heatmap_results} show a strong correlation regardless of the contamination rate employed.

\section{Conclusions}

In this work we explore robust methods based on the Rényi, Tsallis and Kaniadakis generalized statistics. Since the solution of the data-inversion strongly depends on the employed objective function, the generalized objective functions are indeed valuable for this purpose. In fact, given a proper choice of the entropic index, it is possible to get a robust objective function that handles errors that do not obey the Gaussian statistics.

In particular, we investigate a special example of non-Gaussian errors: outliers. In this scenario, we seek to answer some basic questions: (i) what is the most appropriate choice for entropic indexes? (ii) Which of the proposed methods is more resistant to outliers? For the first question, we find that there is a limit for every method in which the objective functions are able to ignore aberrant values without compromising the results. In addition, we note that the properties of the $\alpha$-, $q$- and $\kappa$-generalized distributions and the respective objective functions have similar characteristics at the limit: $(\alpha, \, q, \, \kappa) \rightarrow (1/3, \, 3, \, 2/3)$.

To conclude, it is worth emphasizing that although these methodologies have been successfully employed in geophysical applications, our proposals are easily adaptable to a wide variety of parameter estimation problems. In this regard, we hope that the methodologies proposed in this work are of great value for the modelling of complex systems with numerous unknown variables, as the generalized objective functions are able to reduce the computational cost by accelerating the convergence of the process of optimization, as shown by analysing the influence function.

\section*{Acknowledgements} 
\label{sec:acknowledgements}
J.V.T. de Lima, J.M. de Araújo, G. Corso and G.Z. dos Santos Lima gratefully acknowledge support from \textit{Petrobras} through the  project "\textit{Statistical physics inversion for multi-parameters in reservoir characterisation}" at Federal University of Rio Grande do Norte. J.M. de Araújo thanks \textit{Conselho Nacional de Desenvolvimento Científico e Tecnológico} (\textit{CNPq}) for his productivity fellowship (grant no. 313431/2018-3). G. Corso acknowledges \textit{CNPq} for support through productivity fellowship (grant no. 307907/2019-8).

\section*{Declarations}
\subsection*{Conflicts of interest}
The authors declare that they have no conflict of interest.

\section*{Author contribution statement}
All authors contributed equally to this work.

%Bibliography
\bibliographystyle{unsrt}  
\bibliography{references}

\end{document}